\documentclass[journal]{IEEEtran}
\usepackage{cite}

\ifCLASSINFOpdf
  \usepackage[pdftex]{graphicx}
  \graphicspath{/figures}
  \DeclareGraphicsExtensions{.pdf,.jpeg,.png}
\else
\fi
\usepackage{array}

\ifCLASSOPTIONcompsoc  \usepackage[caption=false,font=normalsize,labelfont=sf,textfont=sf]{subfig}
\else
 \usepackage[caption=false,font=footnotesize]{subfig}
\usepackage{fixltx2e}
\usepackage{booktabs}
\usepackage{amsmath}
\usepackage{tikz}
\usepackage{gensymb}
\usepackage{graphicx}

\usepackage[english, ruled, linesnumbered]{algorithm2e}
\usepackage{algorithmic}

\begin{document}

\tikzstyle{neuron}=[circle, draw=black, minimum size=17pt, inner sep=0pt]
\tikzstyle{input neuron}=[neuron, fill=green!50]
\tikzstyle{hidden neuron}=[neuron, fill=blue!50]
\tikzstyle{output neuron}=[neuron, fill=red!50]
\tikzstyle{annot} = [text width=4em, text centered]

\bstctlcite{bibliography:BSTcontrol}
%
\title{Low Complexity Kolmogorov-Arnold Network-based DPD for Analog RoF Fronthaul}
%
%
%

\author{Carlos Daniel Fontes da Silva, Tianyu Jiang, Lu Zhang, Vjaceslavs Bobrovs, Xianbin Yu, Xiaodan Pang, Oskars Ozolins, \vspace{-0.5cm}
Edson~Porto~da~Silva,~\IEEEmembership{Senior Member,~IEEE}, 

\thanks{Carlos Daniel Fontes da Silva and Edson Porto da Silva are with the Department
of Electrical Engineering, Federal University of Campina Grande (UFCG), Campina Grande, PB, 58429-900, Brazil. (e-mails: carlos.fontes@ee.ufcg.edu.br, edson.silva@dee.ufcg.edu.br)

Tianyu Jiang is with the Department of Applied Physics, KTH Royal Institute of Technology, 106 91 Stockholm, Sweden (e-mail: tianyuj@kth.se).

Lu Zhang and Xianbin Yu are with College of Information Science and Electronic Engineering, Zhejiang University, Hangzhou, China (e-mails: zhanglu1993@zju.edu.cn, xyu@zju.edu.cn).

Xiaodan Pang is with Institute of Telecommunications, Riga Technical University, 1048 Riga, Latvia, and College of Information Science and Electronic Engineering, Zhejiang University, Hangzhou, China (e-mail: xiaodan.pang@ieee.org).

Vjaceslavs Bobrovs and Oskars Ozolins are with the Institute of Telecommunications, Riga Technical University, 1048 Riga, Latvia (e-mail: vjaceslavs.bobrovs@rtu.lv, oskars.ozolins@rtu.lv)}
\thanks{Manuscript received February 26, 2026; revised .}}

%
%

\markboth{Journal of Lightwave Technology, ~Vol.~XX, No.~X, February~2026}%
{Shell \MakeLowercase{\textit{et al.}}: Bare Demo of IEEEtran.cls for IEEE Journals}
%



\maketitle


\begin{abstract}

This paper proposes and demonstrates experimentally for the first time a Kolmogorov–Arnold Network (KAN)–based digital predistortion (DPD) model, named envelope time-delay KAN (ETDKAN), for mitigating nonlinear distortions in analog radio-over-fiber (A-RoF) systems. The ETDKAN model incorporates physical constraints of radio-frequency (RF) nonlinear devices and, through KAN symbolization, achieves a significant reduction in computational complexity while improving interpretability. The proposed model is numerically implemented and optimized alongside multilayer perceptron (MLP) and memory-polynomial-based DPDs. Results show that the resulting symbolic ETDKAN (symbETDKAN) attains ACLR and EVM performance comparable to neural network-based models, while maintaining a computational complexity close to that of memory polynomials. Experimental validation using an A-RoF system confirms the practical feasibility of the proposed approach, which resulted in a 4–5 dB reduction in ACLR in the analyzed scenario.

\end{abstract}


%
\IEEEpeerreviewmaketitle

\section{Introduction}\label{sec_intro}
%
%
%
%
\IEEEPARstart{T}{he} forthcoming generation of mobile wireless communication systems (6G) is expected to comply with formidable requirements for both high data rates and seamless connectivity \cite{alsabah2021}. Expanding the capacity in the physical layer of 6G networks will require the use of higher frequency bands, such as Terahertz (THz) and millimeter Wave (mmW) spectra, along with the development of solutions to overcome numerous challenges inherent to communication across these frequency bands. A major obstacle is the increased levels of attenuation that wireless signals within these bands experience when propagating over the atmosphere \cite{uwaechia2020}. High levels of attenuation reduce transmission reach, thus impacting in the coverage of mobile access networks. 

One potential solution for extending broadband radio signals' reach involves using analog Radio-over-Fiber (A-RoF) systems \cite{lim2021}. A-RoF leverages the high bandwidth and low attenuation of optical fibers to transport broadband wireless signals across significantly greater distances effectively. This is achieved by modulating the amplitude of an optical carrier with the wireless signal, transmitting the resulting modulated carrier over the fiber link, and converting the signal from the optical domain back to the electrical domain through envelope detection (i.e., direct detection). Following detection, a remote antenna can re-transmit the wireless signal to its intended destination. 

However, signals propagating through A-RoF links are subject to impairments that may compromise the quality of transmission (QoT). These impairments may stem from either linear or nonlinear phenomena. In the optical domain, the nonlinear characteristics of electro-optic modulators, such as Mach-Zhender modulators (MZMs), contribute to signal distortion during the optical modulation process. Additionally, linear dispersive effects and nonlinear Kerr effects within the fiber channel can influence the envelope of the optical carrier, with their impact dependent on factors such as the propagation distance and the power sent into the fiber. In the electrical domain, several hardware impairments can also degrade signal quality. These include component bandwidth limitations, in-phase/quadrature (IQ) imbalances, and, notably, the nonlinear behavior of power amplifiers (PAs) employed to boost the power of the radio signals. 

Nonlinear distortions are of critical importance, as they are directly dependent on the power level of the signals, which plays a central role in trade-offs of power efficiency and QoT. Usually, the power efficiency of PAs is maximized when they operate in the so-called saturation regime, where the nonlinear effects of the amplification process are enhanced. Moreover, the optical power at the output of optical modulators depends on the power of the modulating signal. Thus, the signal-to-noise ratio (SNR) at the end of the A-RoF link depends on the power level of the modulating signal at the optical modulator. 

Digital predistortion (DPD) is an effective strategy to mitigate performance degradation caused by nonlinearities within the A-RoF link, particularly the distortions induced by the operation of the MZM and the PA. DPD techniques involve two basic steps: estimating the inverse of the channel's input-to-output nonlinear response, and applying this inverse response to the input signal before transmission through the A-RoF channel, which linearizes the overall system response, minimizing performance degradation due to channel nonlinearities. In PA linearization for wireless systems, models derived from Volterra series, such as the memory polynomials (MP), are well-established in literature \cite{morgan2006}.
\newpage
Recently, due to their ability to approximate intricate input-output nonlinear multidimensional mappings, neural networks (NNs) have been explored in the literature as potential candidates to implement nonlinear predistortion to linearize A-RoF links. These networks can learn the inverse nonlinear response of the A-RoF channel via supervised learning, effectively capturing intricate patterns and distortions introduced by nonlinear components such as PAs and MZMs. Once trained, the NN can be utilized to implement DPD with the objective of linearizing the channel and enhancing transmission performance.

Several authors have discussed NN-based linearization of PAs. In \cite{wang2019}, an augmented real-valued time-delay neural network (ARVTDNN) is proposed to address nonlinear distortions in wideband direct-conversion transmitters. Unlike the real-valued time-delay neural network (RVTDNN), which only uses the IQ components of the baseband signal as input features, the ARVTDNN model includes additional envelope-dependent terms. This approach provides the neural network with extra input features, improving nonlinear distortion modeling and DPD capabilities. Simulations and experiments demonstrate that the ARVTDNN outperforms the RVTDNN, offering reduced complexity and improved numerical stability. 


In \cite{tanio2020}, the envelope time-delay neural network model (ETDNN) is proposed. It incorporates as input elements the time-delayed signal amplitudes and combines the network output with the complex inputs, forming a model that satisfies the parity constraints of PA modeling \cite{lima2009}. When comparing the performance of the DPD based on this model, in terms of transmission metrics and computational complexity, with the ARVTDNN and RVTDNN models in compensating for a PA’s nonlinearities, the ETDNN outperformed them in both aspects in the analysed scenario.

In \cite{liu2025}, a neural network model based on the Kolmogorov–Arnold Representation Theorem is presented. Kolmogorov–Arnold networks (KAN) differ from Multilayer Perceptron (MLP) networks in that they use parameterizable activation functions, which are optimized during the training stage instead of the weights and biases. This model has been investigated as a promising alternative, especially due to its superior interpretability compared to MLPs, allowing for a symbolic representation that can simplify its application. The use of KAN as a DPD model is still little explored; however, \cite{chen2025} proposed the real-valued time-delay KAN (RVTDKAN) model, with a principle similar to that of the RVTDNN, which was successfully applied to compensate for the nonlinearities of a PA.

In this work, we propose and experimentally demonstrate a DPD model based on KAN networks for linearization of A-RoF links including the non-linearities from PA and MZM. The proposed model is named envelope time-delay KAN (ETDKAN), and can be a good candidate for DPD especially due to the symbolization process which can reduce the computational complexity of the model in comparison to other NN-based DPD models. 

The remaining of the paper is divided as follows: in Section~\ref{sec:KAN}, the KAN is introduced and the ETDKAN for DPD is presented in details; in Section~\ref{sec:models}, a detailed description of the A-RoF communication model considered in this work is provided; in Section~\ref{sec:numerical_results}, the numerical results obtained are discussed; in Section~\ref{sec:experimental_results},
an experimental validation is presented to demonstrate practical feasibility and to confirm consistency with the simulated trends, leading to the conclusions. 

\vspace{-0.25cm}
\section{Kolmorogorov-Arnold Networks}\label{sec:KAN}

KAN networks are based on the Kolmogorov–Arnold Representation Theorem, which states that a multivariable function $f(x_1,\cdots,x_n)$ can be represented as a composition of univariate functions as follows \cite{liu2025}:

\begin{equation}\label{eq:KA_theorem}
    f(x_1, \cdots, x_n) = \sum_{q = 1}^{2n+1} \Phi_{q}\left( \sum_{p=1}^n \phi_{q, p}(x_p) \right)
\end{equation}

\begin{figure}[!hbt]
    \centering
    \includegraphics[width=\linewidth]{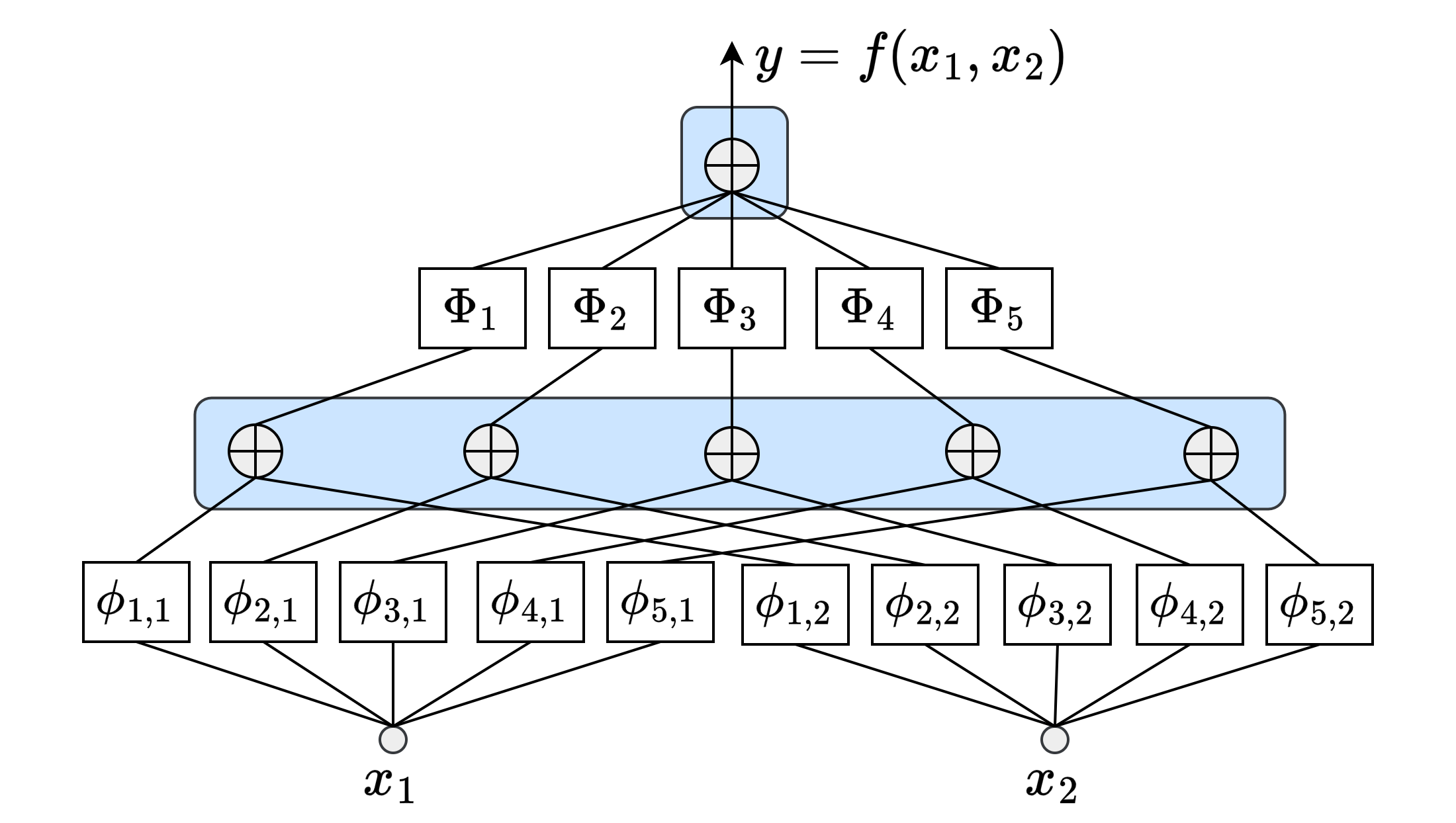}
    \caption{KAN diagram of a network representing a function $f(x_1,x_2)$ as composition of univariate functions.}
    \label{fig:KAN_example}
\end{figure}

For $n=2$, \eqref{eq:KA_theorem} can be represented by a diagram such as in Fig. \ref{fig:KAN_example}, similar to a MLP network with one hidden layer, but instead of neurons, the nodes in each layer are adders that take as input the output of the activation functions. The principle of KAN networks consists in generalizing \eqref{eq:KA_theorem}, assuming that more layers can be included and that the number of nodes per layer is a free parameter.

Therefore, consider a KAN network with $L+1$ layers, where each layer has $n_l$ nodes and activation functions $\phi^{(l)}_{i,j}$, $1 \leq i \leq n_{l+1}$, $1 \leq j \leq n_{l}$, $0\leq l \leq L$. The relationship between the vectors of two consecutive layers is given by:

\begin{equation}\label{eq:rel_vetores_kan}
    \mathbf{x}^{(l)} = 
    \begin{bmatrix}
        \displaystyle\sum_{j=1}^{n_{l-1}} \phi_{1,j}^{(l-1)}\left( x_{j}^{(l-1)} \right) \\
        \displaystyle\sum_{j=1}^{n_{l-1}} \phi_{2,j}^{(l-1)}\left( x_{j}^{(l-1)} \right) \\ 
        \vdots \\ 
        \displaystyle\sum_{j=1}^{n_{l-1}} \phi_{n_l,j}^{(l-1)}\left( x_{j}^{(l-1)} \right)
    \end{bmatrix}
\end{equation}

As in MLP networks, training KAN networks consists of determining a set of parameters that minimize an objective function, and in the case of KAN networks, these parameters are associated with the activation functions. In \cite{liu2025}, each activation function $\phi(x)$ is represented as the sum of a fixed nonlinear function $b(x)$ (in this case, $b(x) = \text{SiLU}(x)$) and a B-spline $S_k(x)$ of degree $k$, weighted by the coefficients $w_b$ and $w_s$.

\begin{equation}\label{eq:func_ativacao_kan}
    \phi(x) = w_b b(x) + w_s S_k(x)
\end{equation}

B-splines are parametric functions obtained from the combination of basis functions $B_{i,k}(x)$ weighted by $P$ control points $c_i$, as presented in \eqref{eq:bspline_soma}.

\begin{equation}\label{eq:bspline_soma}
    S_k(x) = \sum_{i} c_i B_{i,k}(x)
\end{equation}

Each basis function $B_{i,k}(x)$ is represented by polynomials of degree $k$ defined on subintervals established by a knot vector $\mathbf{T} = [ t_0, t_1, \cdots, t_m ]$, where $t_0 \leq t_1 \leq \cdots \leq t_m$, with $m = P+k$. The expressions for $B_{i,k}(x)$ can be derived using the recursive Cox-de Boor algorithm:

\begin{align}
    B_{i,0}(x) &= 
    \begin{cases}
        1\ ,\ \text{se } t_{i} \leq x \leq t_{i+1} \\
        0\ ,\ \text{c.c.}
    \end{cases} \\
    B_{i,k}(x) &= \dfrac{x - t_i}{t_{i+k} - t_i} B_{i,k-1}(x) + \dfrac{t_{i+k+1} - x}{t_{i+k+1} - t_{i+1}} B_{i+1,k-1}(x)
\end{align}

\subsection{ETDKAN}

\begin{figure}[!th]
    \centering
    \includegraphics[width=0.95\linewidth]{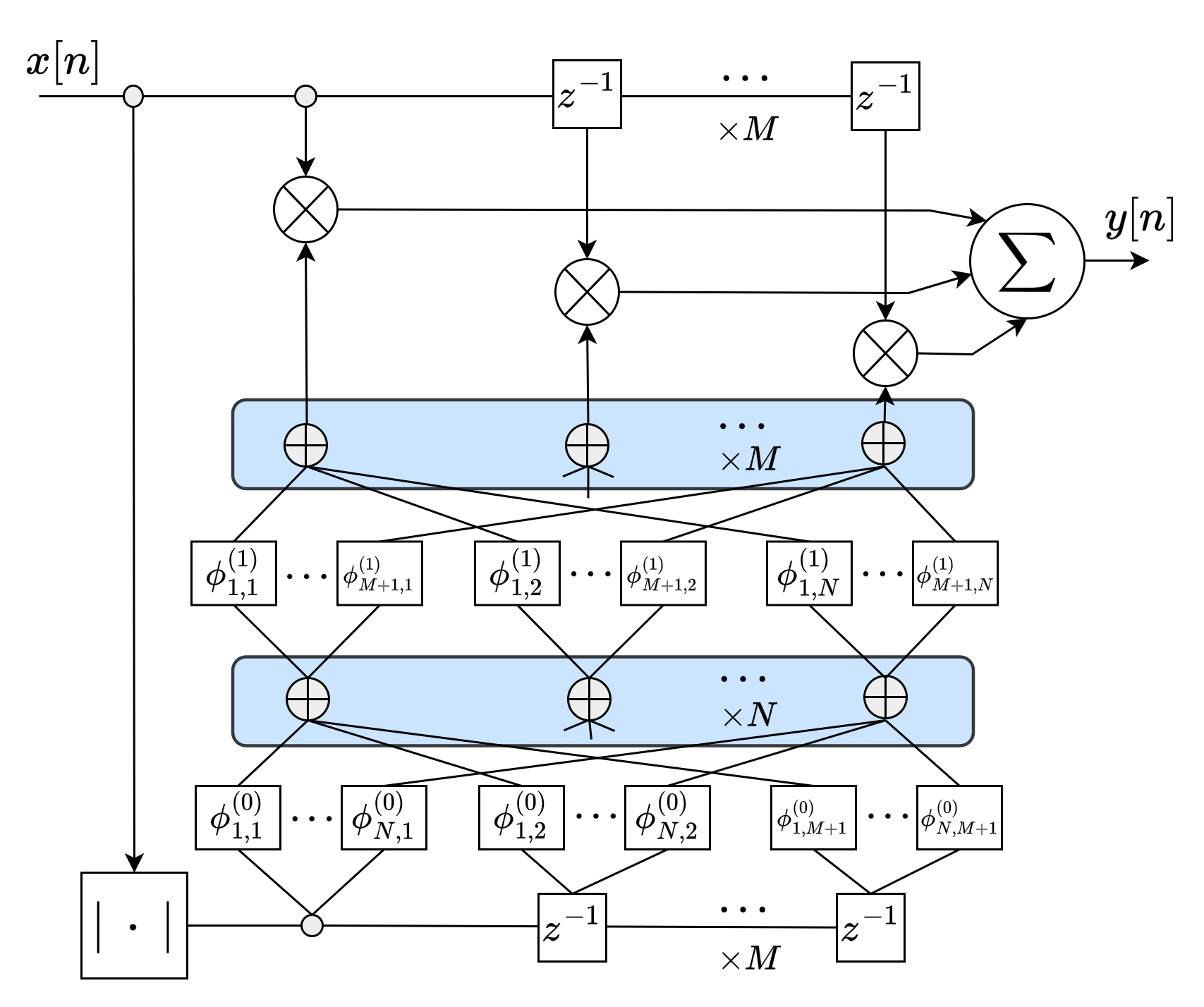}
    \caption{Diagram of proposed DPD based on envelope time-delay Kolmogorov- -Arnold network.}
    \label{fig:ETDKAN}
\end{figure}

The proposed ETDKAN model consists of an adaptation of the ETDNN \cite{tanio2020} in which the MLP is replaced by a KAN with $M$ inputs and $M$ outputs, and a hidden layer of size $N$. The input to the KAN network consists of a sequence of $M$ delays of the magnitude of the input signal $x[n]$, and $y[n]$ is a linear combination of $x[n]$ with the output of the internal KAN network, as illustrated in Fig. \ref{fig:ETDKAN}.

\begin{equation}\label{eq:out_ETDKAN}
    y[n]= \sum_{m=0}^{M} \left( \sum_{i=1}^{N} \phi^{(1)}_{m+1,i} \left( \sum_{j=0}^{M} \phi^{(0)}_{i,j+1} \big(\big|x[n-j]\big|\big) \right) \right) x[n-m]
\end{equation}


From \eqref{eq:out_ETDKAN}, it can be seen that the output $y[n]$ is an odd function of the input $x[n]$, a condition that is necessary for satisfying the physical constraints of PA modeling and its inverse response \cite{lima2009}. This condition is not necessarily satisfied by ARVTDNN and RVTDNN models, which may result in performance losses, as suggested in \cite{tanio2020}.

\subsection{Symbolic KAN}
\begin{figure}[!b]
    \centering
    \includegraphics[width=0.95\linewidth]{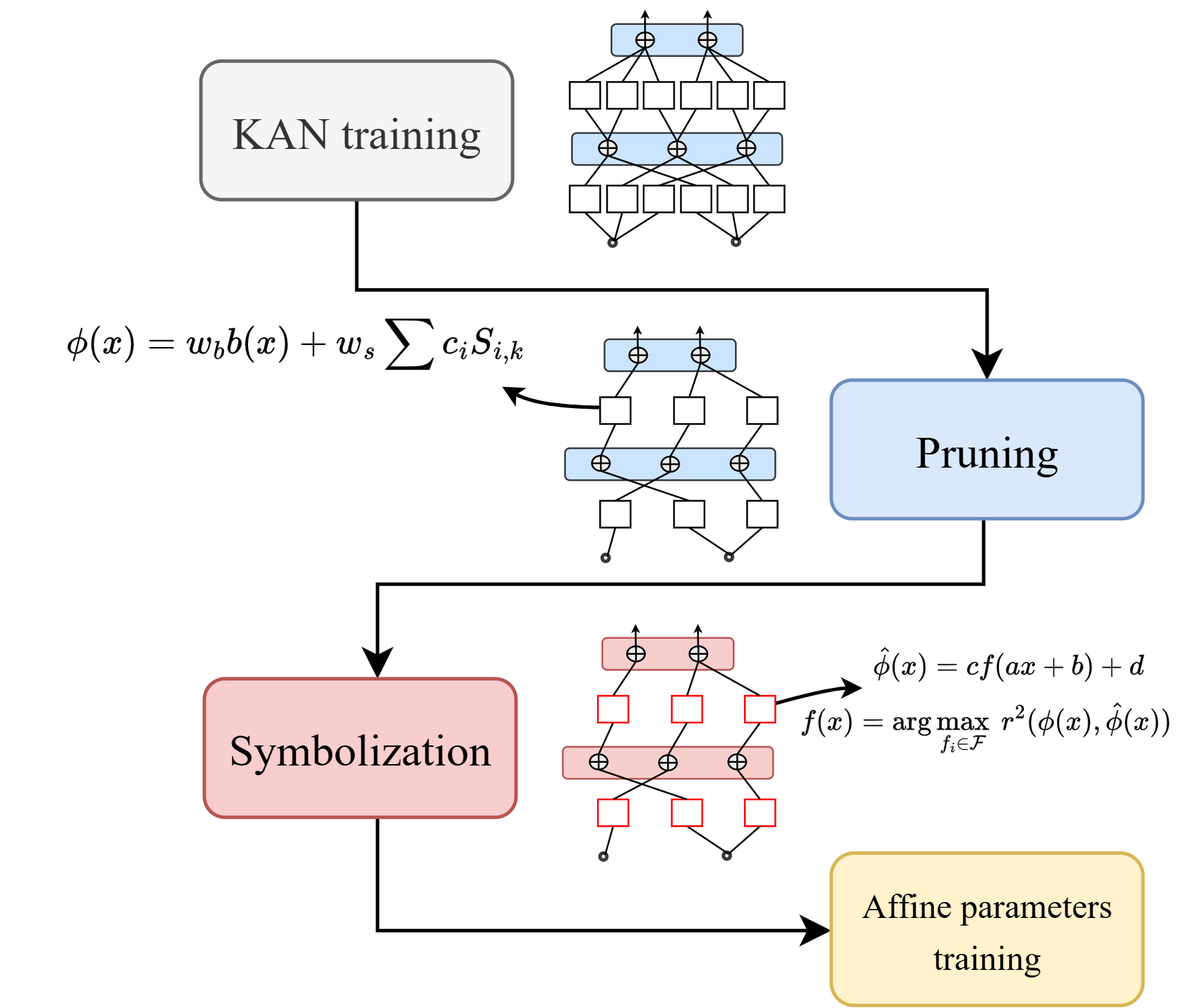}
    \caption{Diagram of KAN training with symbolization.}
    \label{fig:KAN_training_diagram}
\end{figure}

During the training of the KAN network, the optimizable parameters in each activation function are the weights $w_s$ and $w_b$, and the B-splines' control points. Once the models are trained, they can be pruned, and the activation functions can then be approximated by analytical functions using symbolic regression, which can reduce the computational cost of the model. This process is proposed by \cite{liu2025}, and it is illustrated in Figure \ref{fig:KAN_training_diagram}.

In the symbolic model, an activation function $\phi(x)$ is approximated by an expression of the form $\hat{\phi}(x) = cf(ax+b) + d$, where $a$, $b$, $c$, and $d$ are the affine parameters, and $f(\cdot)$ an analytical function. Initially, the values $c = 1$ and $d = 0$ are fixed, and through a grid search, the values of $a$ and $b$ that maximize the coefficient of determination $r^2\big(\phi(x); f(ax+b)\big)$, defined in \eqref{eq:r2}, are selected.

\begin{equation}\label{eq:r2}
    r^2(\alpha;\beta) = \dfrac{\left(\displaystyle\sum_i \left(  \alpha(x_i) - \overline{\alpha(x_i)} \right) \left(  \beta(x_i) - \overline{\beta(x_i)} \right)\right)^2 }
    {\displaystyle\sum_i \Big( \alpha(x_i) - \overline{\alpha}(x_i)  \Big)^2 \displaystyle\sum_i \Big( \beta(x_i) - \overline{\beta(x_i)}  \Big)^2}
\end{equation}

Once the parameters $a$ and $b$ are defined, $c$ and $d$ are computed via linear regression. This process must be repeated for a set of candidate functions $\mathcal{F} = \{ f_1, f_2, \cdots, f_i, \cdots \}$, so that the selected function is the one with the highest $r^2$. After symbolizing all activation functions in the network, the training can proceed to ensure greater accuracy of the affine parameters of each symbolic function.

\vspace{-0.25cm}
\section{DPD for A-RoF Systems}\label{sec:models}

Due to its easy reconfigurability and effective linearization capability, digital predistortion is a well-established technique for compensating nonlinearities of PAs in wireless systems \cite{leduc2020}. More recently, it has attracted significant interest for application in A-RoF systems \cite{pereira2023}, where nonlinear distortions arise not only from the PA but also from the optical modulation stage. In DPD, an estimate of the inverse nonlinear response of the channel is applied to the baseband signal prior to transmission. The distortions introduced by the DPD block are then cancelled by the channel, thereby linearizing the overall system response.

There are two main approaches to obtaining the inverse response of the channel: the Direct Learning Architecture (DLA) and the Indirect Learning Architecture (ILA). In DLA, the DPD model is trained in its operating position as a predistorter, which requires a channel model to be defined and included in the training loop. In contrast, in ILA the DPD block is temporarily placed after the channel and trained as a postdistorter. Using the channel output as the model input and the transmitted signal as the target, the DPD parameters are optimized. Once the model converges, it is then repositioned to operate as a predistorter.

\begin{figure}[!htb]
    \centering
    \includegraphics[width=\linewidth]{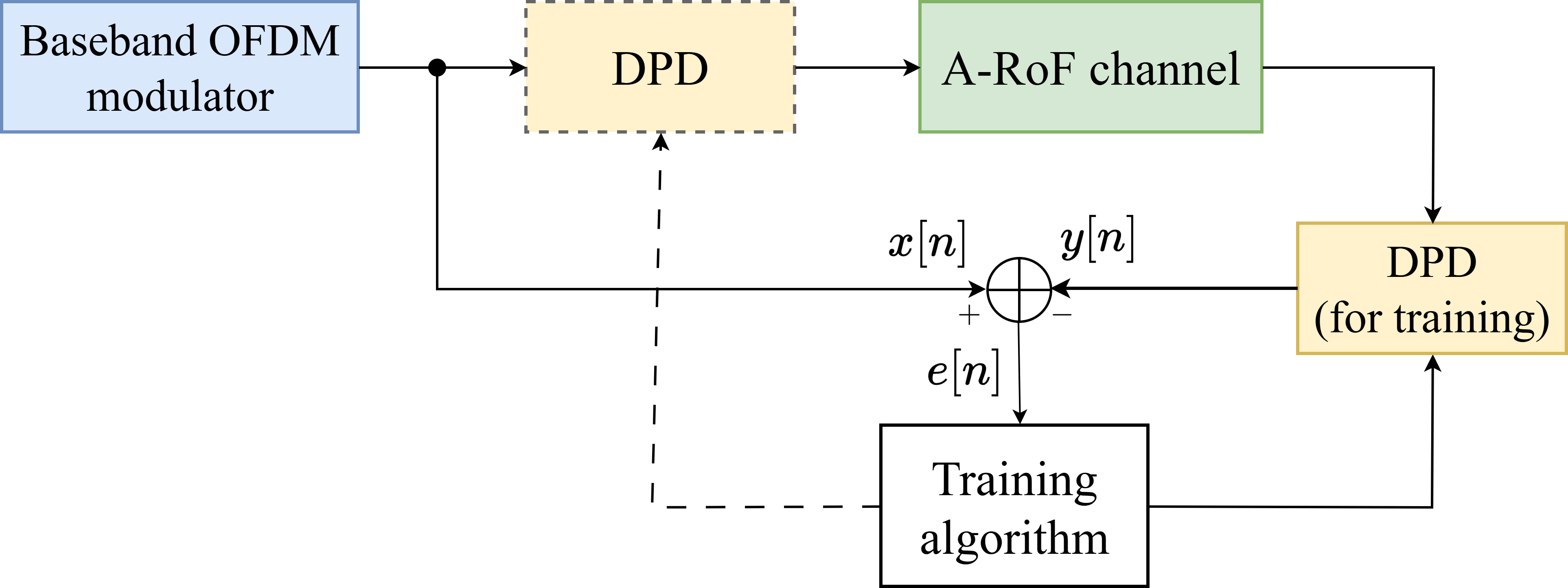}
    \caption{Indirect Learning Architecture (ILA) for DPD training.}
    \label{fig:ILA}
\end{figure}

In the following subsections, the numerical model implemented of the A-RoF system and its nonlinearities are explained in details.

\begin{figure*}[t!]
    \centering
    \includegraphics[width=0.95\linewidth]{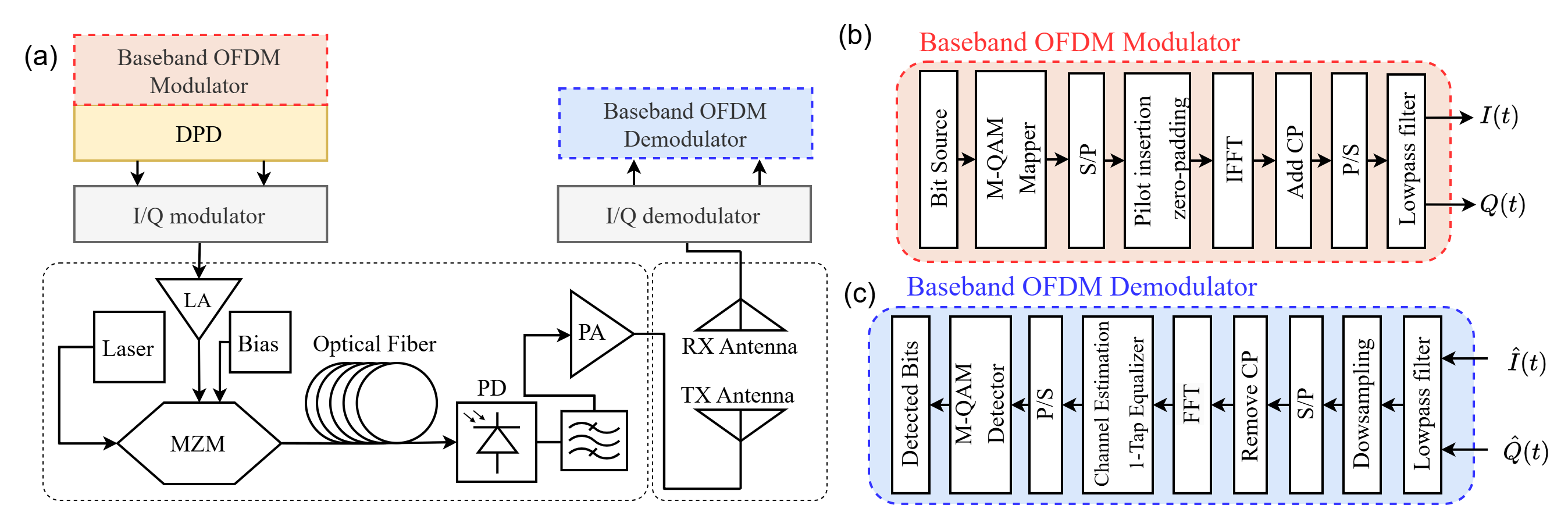}
    \caption{System model considered in this paper. (a) Block diagram for the A-RoF transmission system model; (b) Signal processing assumed for the OFDM baseband modulator; (c) Signal processing assumed for the OFDM baseband demodulator.}
    \label{fig:simulation_setup}
\end{figure*}

\vspace{-0.25cm}
\subsection{Waveform generation}
Due to its widespread use in mobile communications, orthogonal frequency division multiplexing (OFDM) is considered here as the type of waveform to be transported through the A-RoF channel \cite{alsabah2021}. In Fig.~\ref{fig:simulation_setup} (b)-(c) the setup of the OFDM baseband modulator and the OFDM baseband demodulator are detailed. The channel amplitude and phase response are estimated from the pilot symbols via linear interpolation. The estimated channel frequency response is used to perform zero-forcing 1-tap equalization of the received signal. Finally, the received constellation symbols can be used to evaluate the QoT. 

A summary of the OFDM signal parameters is presented in Table~\ref{tab:ofdm_parameters}. The signal has a bandwidth ($B_w$) of 100 MHz, includes a subcarrier spacing of 30 kHz and 273 resource blocks (RBs). The chosen of 3.55 GHz for electrical carrier frequency is due to its wildly implements in current 5G deployments and still has a considerable potential in bandwidth.

\begin{table}[htbp]
    \centering
    \caption{OFDM signal parameters}
    \label{tab:ofdm_parameters}
    \begin{tabular}{@{}ll@{}}
    \toprule
    \textbf{Parameter} & \textbf{Value} \\ \midrule
    Number of subcarriers ($N$)                  & 4096  \\
    Number of information subcarriers ($N_{s}$)  & 3243   \\
    Number of pilot subcarriers ($N_{p}$)        & 32 \\
    Number of null subcarriers ($N_{z}$)         & 821 \\
    Subcarrier spacing ($\Delta f$)              & $\approx$ 30~kHz        \\
    Bandwidth ($B_w$)                            & 100~MHz   \\
    Cyclic prefix length ($N_{\text{CP}}$)       & 288   \\
    Adjacent channel leakage ratio (ACLR)        & -44.70 dB \\
    RF center frequency                          & 3.55~GHz \\
    Simulation sampling rate ($f_s$)             & 12.5 GSa/s        \\   
    Modulation scheme                            & 64QAM  \\
    Channel estimation                           & linear interpolation    \\ \bottomrule
    \end{tabular}
\end{table}

\begin{table}[!b]
    \centering
    \caption{Optical channel parameters}
    \label{tab:channel_parameters}
    \begin{tabular}{@{}ll@{}}
    \toprule
    \textbf{Parameter}    & \textbf{Value}     \\ \midrule
    MZM's $V_{\pi}$       & 3~V           \\
    MZM's optical input power  & 15~dBm   \\
    Fiber length         & 25~km \\
    Fiber dispersion coefficient         & 16~ps/nm/km \\
    Fiber attenuation     & 0.2~dB/km        \\
    Photodiode bandwidth   & 12~GHz \\
    Photodiode temperature   & 25~C$^{\degree}$   \\
    Photodiode dark current &  5~nA  \\ \bottomrule   
    \end{tabular}
\end{table}

\vspace{-0.25cm}
\subsection{Optical Modulator Model}

The MZM model is defined under the assumption that the device is constructed in a push-pull configuration \cite{seimetz2009}. The input-output relation of the MZM is defined in eqs.~(\ref{MZM1})-(\ref{MZM2}),

\begin{equation}\label{MZM1}
    |E_{out}(t)|=|E_{in}(t)\cos \varphi(t)|
\end{equation}
\begin{equation}\label{MZM2}
    \varphi(t)=\frac{\pi}{2V_\pi}\left[u(t)+V_b\right] 
\end{equation}


\noindent where $E_{in}(t)$ is the envelope of the input optical field, $E_{out}(t)$ is the envelope of the output optical field, $u(t)$ is the modulating signal, $V_\pi$ is the half-wave voltage of the MZM, and $V_b$ is the DC bias added to $u(t)$, set as $-V_{\pi}/2$ such that the MZM is operated at the quadrature point. 
\vspace{-0.25cm}
\subsection{Optical Fiber Model}

In this paper, a linear model is assumed for the fiber channel, accounting for loss and chromatic dispersion (CD). In direct-detection systems the combination of CD and the square law detection of photodiodes introduces frequency-selective power fading that may cause severe penalties to the transmission performance. Consequently, signals propagating over fiber links in A-RoF systems may experience frequency-selective power fading whose intensity depends on the RF carrier frequency and propagation distance, leading to performance penalties. The baseband equivalent frequency response $H_{\mathrm{CD}}(f)$ of CD  after square law detection is approximated by \cite{rizzelli2024}

\begin{equation}\label{CDtransfer}
    H_{\mathrm{CD}}(f)=\cos \left((2 \pi f)^2 \beta_2l / 2\right)
\end{equation}

\noindent where $\beta_2=-D \lambda_c^2 / 2 \pi c$, $D$ is the fiber dispersion parameter, $\lambda_c$ is the wavelength of the optical carrier, and $c$ is the speed of light. In this work, the carrier frequency of the RF signal is not high enough for this effect to have an impact, so the optical link response is essentially flat over the frequency range used.

\vspace{-0.25cm}
\subsection{Photodiode Model}

After propagating through the fiber, the optical signal is sent to a photodiode where the RF signal modulating the optical carrier is downconverted to its original frequency.
The model assumed for the photodiode considers the presence of thermal and shot noise \cite{seimetz2009}, as well as a bandwidth limitation. It is assumed that the photocurrent follows an ideal square-law detection with no saturation effects. A summary of the optical channel parameters is presented in Table~\ref{tab:channel_parameters}.

\vspace{-0.25cm}
\subsection{Power Amplifier Model}

After traversing the A-RoF link and before reaching the remote transmit antenna, the signals are amplified by a PA, where both their phase and amplitude are affected by nonlinear distortions. One commonly employed method to characterize the nonlinearity introduced by the PA involves utilizing an amplitude modulation-amplitude modulation (AM-AM) model and an amplitude modulation-phase modulation (AM-PM) model \cite{morgan2006}. In this study, we adopt the memoryless AM-AM AM-PM nonlinear PA model specified by the 3GPP TSG-RAN WG4 task group \cite{Nokia2016}.

Let $s(t) = |s(t)|e^{j\phi(t)}$ be the equivalent baseband complex-valued envelope of the signal at the input of the PA. The baseband AM-AM model is described as 
\begin{equation}\label{AM-AM-PA-model}
A(|s(t)|)=\frac{g |s(t)|}{\left[1+\left(g|s(t)|/v_{sat} \right)^{2\sigma_p}\right]^{1/(2\sigma_p)}},
\end{equation}

\noindent with the typical parameters $g=16$, $\sigma_p=1.1$, $v_{sat}=1.9$~V.

\begin{figure}[!htb]
\centering
\includegraphics[width=1\linewidth]{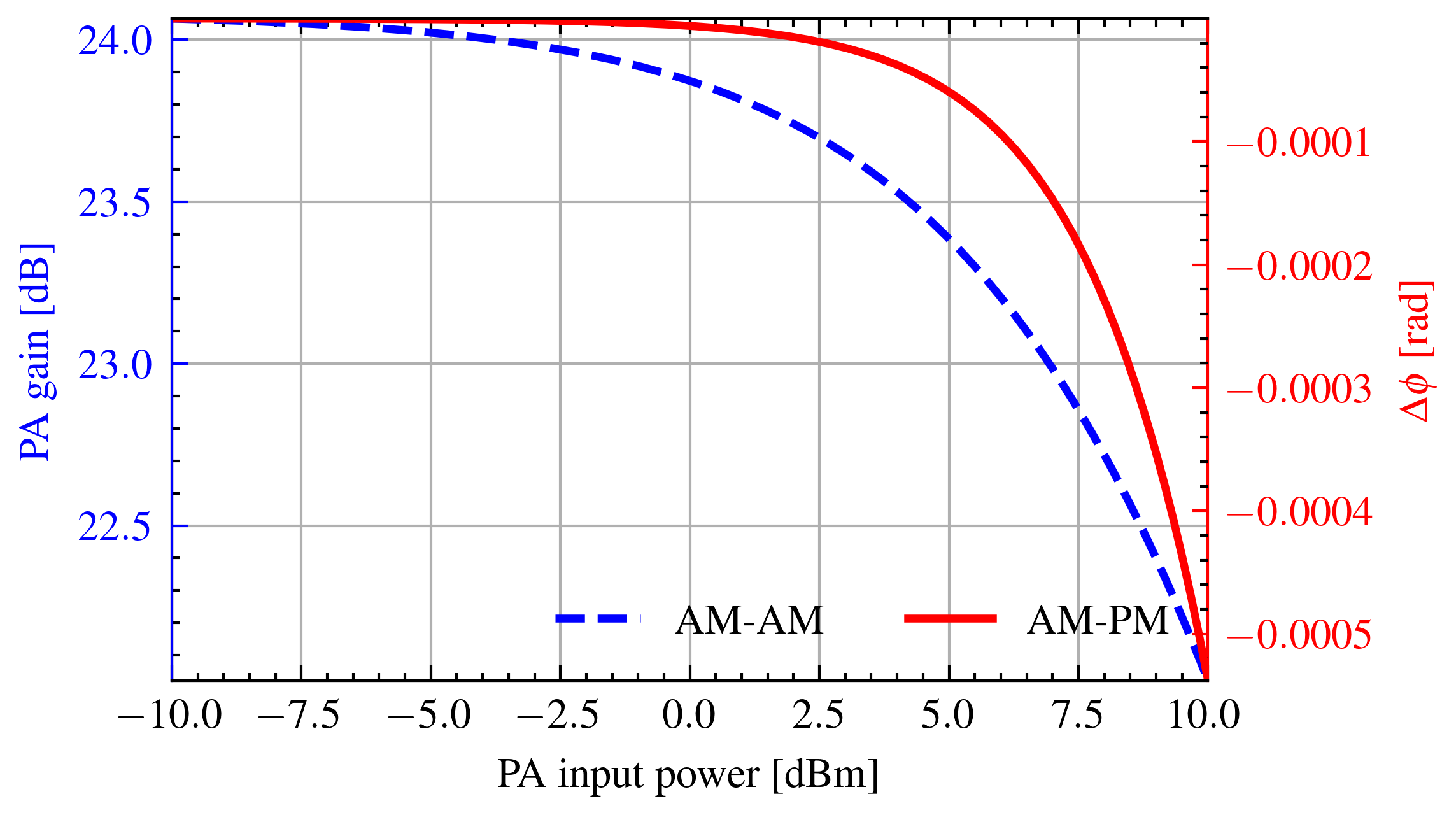}
\caption{Gain and phase shift characteristic curves of PA as a function of input power for the model described in eqs.~(\ref{AM-AM-PA-model})-(\ref{AM-PM-PA-model}). }
\label{fig:pa_transfer_fun}
\end{figure}

The baseband AM-PM model is given by
\begin{equation}\label{AM-PM-PA-model}
\Delta\phi\left(|s(t)|\right)=\frac{\pi}{180}\frac{\alpha |s(t)|^{q}}{\left[1+\left(|s(t)| / \beta\right)^{q}\right]},
\end{equation}

\noindent with parameters $\alpha=-345$, $\beta=0.17$, $q=4$. The gain and the nonlinear phase shift (in radians) induced by the PA model as a function of its input power in dBm are shown in Fig.~\ref{fig:pa_transfer_fun}.

The equivalent baseband signal $\hat{s}(t)$ at the output of the PA can be written as
\begin{equation}
    \hat{s}(t) = A(|s(t)|) e^{j[\phi(t)+\Delta\phi\left(|s(t)|\right)]},
\end{equation}
\noindent which is the resulting signal after $s(t)$ is affected by amplitude and phase nonlinear distortions. Here, it is assumed that the input power to the PA can be controlled independently of the optical power that reaches the photodiode. 

\vspace{-0.25cm}
\section{Numerical results}\label{sec:numerical_results}

Based on the numerically implemented A-RoF channel model, to ensure operation in the nonlinear regime, the RF signal power at the input of the MZM is fixed at 17 dBm, and the signal power at the input of the PA is set to 5 dBm. Under these conditions, the DPD model was trained following the ILA approach, illustrated in Fig. \ref{fig:ILA}.

\begin{figure*}[!t]
    \centering
    \includegraphics[width=1\linewidth]{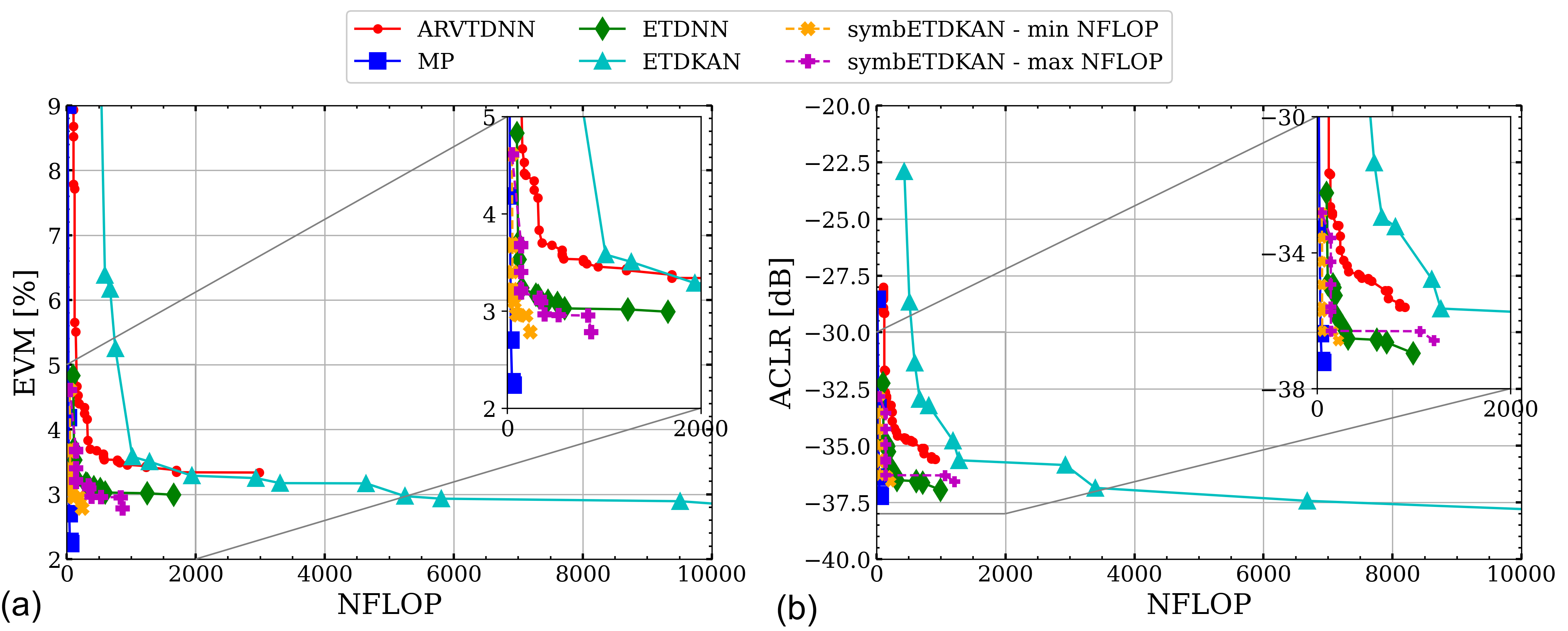}
    \caption{Pareto fronts from multi-objective optimization of DPD's hyperparameters for transmission metrics and complexity.
    (a) EVM vs NFLOP; (b) ACLR vs NFLOP.}
    \label{fig:pareto_fronts}
\end{figure*}

For the performance evaluation of the DPD-ETDKAN, models based on ARVTDNN and ETDNN were also implemented, along with a model based on memory polynomials. For a proper comparison between the models, a multi-objective optimization of each model’s hyperparameters was carried out using the Optuna \cite{optuna} framework. A total of 500 hyperparameter combinations were generated, and for each combination a model was trained. After training, its transmission metrics were computed, as well as the number of floating-point operations (NFLOP) required to compute one output of the model.

To train the NN-based models, arrays containing a total of 50000 signal samples are divided into training and validation datasets in a 75/25 ratio. Training is conducted with a batch size of 1000, using backpropagation and the Adam optimization algorithm with a learning rate of $\mu = 0.005$. For ETDNN and ARVTDNN, 500 epochs were used in the training phase, which was enough to achieve convergence in all configurations; for ETDKAN, it was necessary 100 epochs for convergence. All simulation models and NNs reported in this work were implemented using the Pytorch \cite{pytorch} and pyKAN \cite{liu2025} frameworks. The MP model was trained with the same set of samples, and the Recursive Least Squares (RLS) algorithm adapted to the MP structure was applied \cite{leduc2020}.

For the ETDKAN, the optimization varied the hyperparameters $N$ from 1 to 5 and $M$ from 2 to 8, as well as the order $k$ (from 2 to 6) of the polynomials that compose the B-spline functions, together with a parameter related to the grid size of these functions, all of which influence the model’s accuracy. For the ARVTDNN, the number of hidden layers was varied from 1 to 2, and the number of neurons per layer from 5 to 50, the nonlinearity order was varied from 1 to 5, and the number of memory taps from 2 to 10. For the ETDNN, the number of neurons in the hidden layer was varied from 5 to 50, and those in the input and output layers from 2 to 20. In the memory polynomial model, the number of memory taps was varied from 1 to 10, and the nonlinearity order from 1 to 10. For the MLP-based models, the activation function used in the internal layers was $\text{ReLU}(x)$, and a linear operation was assumed for the output layer.

An alternative model with symbolization, symbETDKAN, was also applied. After the first 50 epochs of training, the model is symbolized, and then more 50 epochs are runned to adjust the affine parameters of symbolic functions. The same hyperparameter optimization range of ETDKAN was applied to this version.

To evaluate the spectral broadening due to nonlinear distortions, the adjacent channel leakage ratio (ACLR) was calculated, which is defined as \eqref{eq:aclr}, where $B_{in}$ and $B_{out}$ refer to the frequencies inside and outside the transmitted OFDM signal bandwidth, respectively.
\begin{equation}\label{eq:aclr}
    \operatorname{ACLR}_\mathrm{dB}=10 \log_{10} \left[\frac{\int_{f \in B_{out}} S(f) d f}{\int_{f \in B_{in}} S(f) d f}\right]
\end{equation}

The other transmission metric computed was the error vector magnitude (EVM), which indicates the degree of dispersion of the received constellation symbols $\hat{X}_r$ relative to the transmitted ones $X_r$.
\begin{equation}\label{eq:def_evm}
    \text{EVM [\%]} = 100\sqrt{\dfrac{ \displaystyle\sum_r \left| \hat{X_r} - X_r \right|^2}{\displaystyle\sum_r \left| X_r \right|^2}}
\end{equation}

For the calculation of the DPD-ETDKAN’s NFLOP, an analytical derivation was performed based on the expressions presented in \cite{yu2024}. The NFLOP for ARVTDNN, ETDNN, and MP was derived analytically by counting the number of additions and real multiplications in a single forward step. Since the symbETDKAN is composed of analytical functions, the NFLOP required for its computation depends on the hardware and software implementation of each function. Therefore, in this work, we assume that the NFLOP ranges from 10 to 100 for each transcendental function in the model, then, for this model, there will be two Pareto fronts: one for the minimum NFLOP and another for the maximum.
\begin{figure}[!t]
    \centering
    \includegraphics[width=0.85\linewidth]{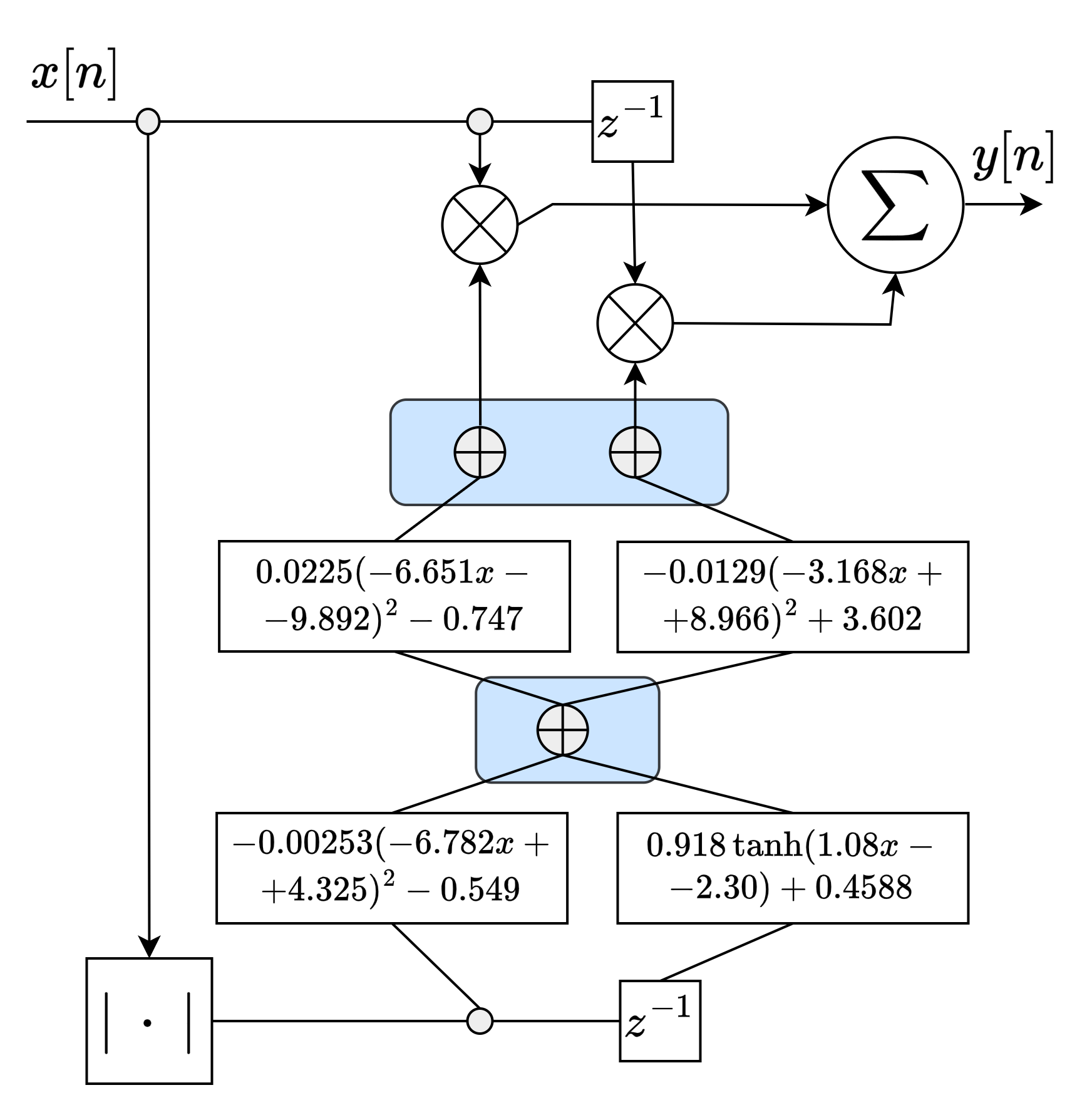}
    \caption{Symbolic model obtained for symbETDKAN-DPD.}
    \label{fig:ETDKAN_symb}
\end{figure}

After hyperparameter optimization, the Pareto curves for the DPD models were obtained, as illustrated in Fig. \ref{fig:pareto_fronts}. It can be observed that, in the analyzed scenario, the MP-DPD achieved superior results in terms of EVM and ACLR with a computational complexity significantly lower than that of the other neural network-based models. The envelope networks outperformed the ARVTDNN, reproducing what was also shown in \cite{tanio2020}. Among the neural network-based models, the ETDKAN-DPD was the only one capable of achieving performance close to that of the MP-DPD in terms of EVM and ACLR, however, this comes at a high computational cost, which is largely due to the computation of the B-splines that make up the KAN network inside the model. The symbETDKAN Pareto fronts reached ACLR and EVM levels comparable to the other NN-based DPD models, but requiring overall lower computational complexity. In fact, it is the model whose optimal points lie closest to the MP model’s complexity. 

\begin{table}[!b]
    \centering
    \caption{Performance metrics of the selected DPD models.}
    \begin{tabular}{|c | c | c | c|}
        \hline
        \textbf{DPD model} & \textbf{NFLOP} & \textbf{EVM [\%]} & \textbf{ACLR [dB]} \\   \hline
        No DPD & -- & 9.13 & -28.26 \\ \hline
        MP & 60 & 2.24 & -37.22 \\ \hline
        ARVTDNN & 852 & 3.54 & -35.48 \\ \hline
        ETDNN  & 620 & 3.05 & -35.62 \\ \hline
        ETDKAN & 5804 & 3.16 & -36.54 \\ \hline
        symbETDKAN & 53 -- 143 & 3.46 & -35.71 \\ \hline
    \end{tabular}
    \label{tab:dpd_metrics}
\end{table}

By considering the ACLR, EVM, and NFLOP metrics, one representative model from each DPD family was selected, and their results are reported in Tab. \ref{tab:dpd_metrics}. The MP-DPD outperformed the NN models in both computacional complexity, requiring fewer NFLOP, and in transmission metrics. The selected symbETDKAN model, illustrated in Fig.\ref{fig:ETDKAN_symb}, is the only one that approaches the NFLOP count of the MP model, with a range from 53 to 143 NFLOP, while achieving transmission metrics similar to the NN-based models. Compared with the ARVTDNN, the symbETDKAN required on average 8 times fewer NFLOPs to achieve similar EVM and ACLR performance, 6 times fewer when compared with the ETDNN, and 59 times fewer compared with the ETDKAN, even with a 0.83 dB difference in ACLR.

\vspace{-0.25cm}
\section{Experimental results}\label{sec:experimental_results}

\begin{figure}[!b]
    \centering
    \includegraphics[width = \linewidth]{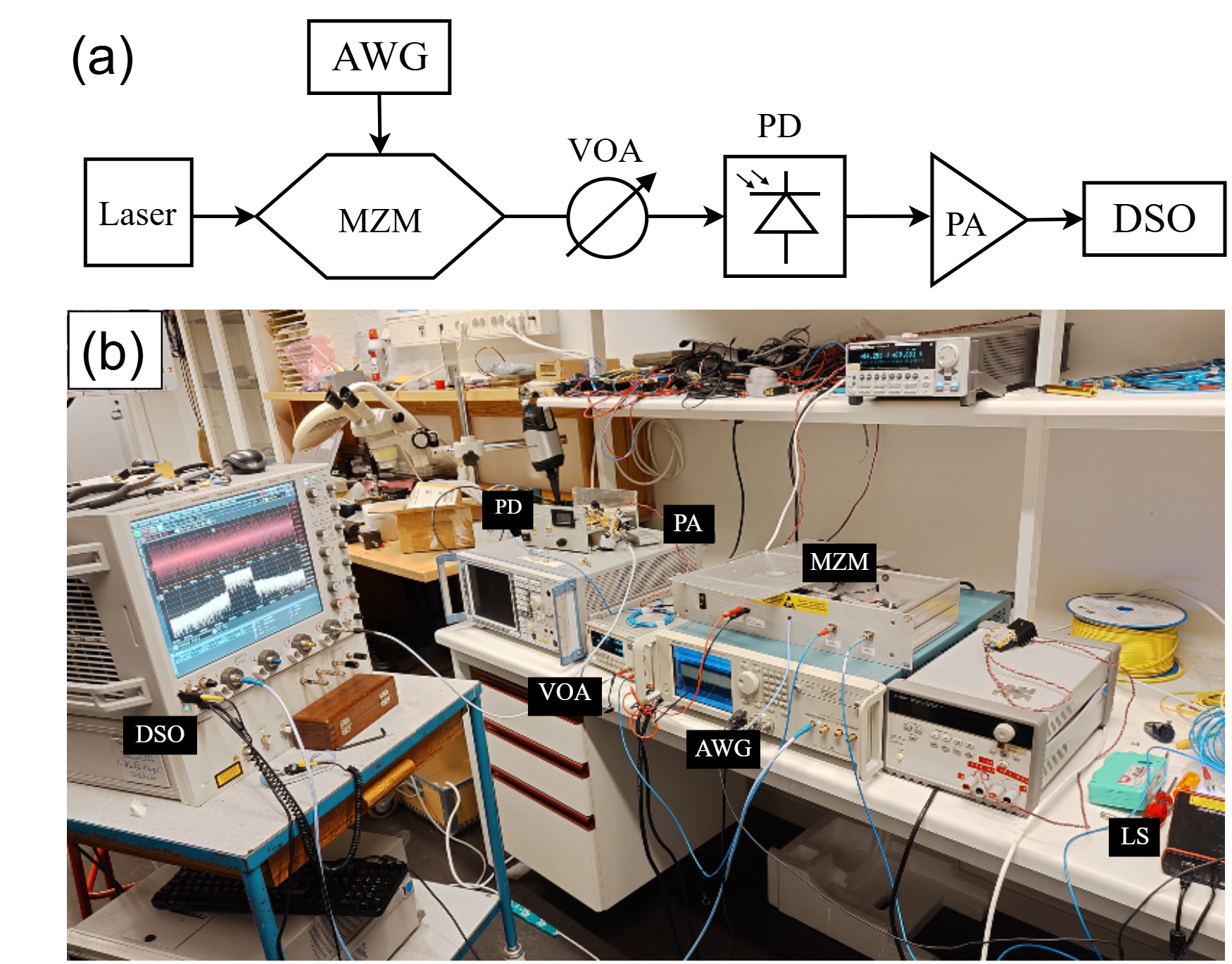}
    \caption{(a) Block diagram and (b) photo of experimental setup used for DPD tests.}
    \label{fig:experimental_setup}
\end{figure}

To access the applicability of the proposed models in a practical environment, an experiment was conducted using a back-to-back A-RoF system, as illustrated in Fig. \ref{fig:experimental_setup}. The light source (LS) is an external cavity laser (ECL) that works at the wavelength of 1550.9 nm, and has a linewidth less than 100 kHz and output the light power at 16 dBm in the experiment. Our test signal is generated by an arbitrary waveform generator (AWG), the sampling frequency is set at 25 GSa/s, with a vertical resolution of 9 bits for the output waveform and an additional 1-bit marker channel for the stability. The signal was generated based on Table \ref{tab:ofdm_parameters} specifications, and it was modulated to the light from LS through a MZM with a $V_{\pi}$ value of 4.2 V. The A-RoF signal output from MZM is controlled by an optical variable attenuator (VOA), and after transmission, the A-RoF signal is received by a photodetector (PD) of 9 GHz bandwidth and 0.8 A/W responsivity. A power amplifier with a linear gain of 22 dB is used to amplify the received signal, which has a noise figure of 6 dB.

After the PA, the signal is captured by a digital storage oscilloscope (DSO) with a sampling frequency set to 20 GSa/s and then sent to a computer for digital signal processing (DSP) operations, such as filtering and synchronization. After processing the received signal, the transmitted and received sequences are used to train the implemented DPD models, following the same procedure presented in the previous section. The signal with DPD is then generated and transmitted again to verify the resulting changes in performance metrics.

Table \ref{tab:dpd_metrics_exp} presents the ACLR, EVM, and NFLOP results for the applied models. Although absolute metric values differ due differences with numerical model, the relative performance ranking among models remains consistent with the simulation results, and the symbETDKAN model also achieves EVM and ACLR values comparable to those of the neural network–based models, while exhibiting computational complexity in terms of NFLOP close to that of the MP model. 

\begin{figure}[!h]
    \centering
    \includegraphics[width = 1\linewidth]{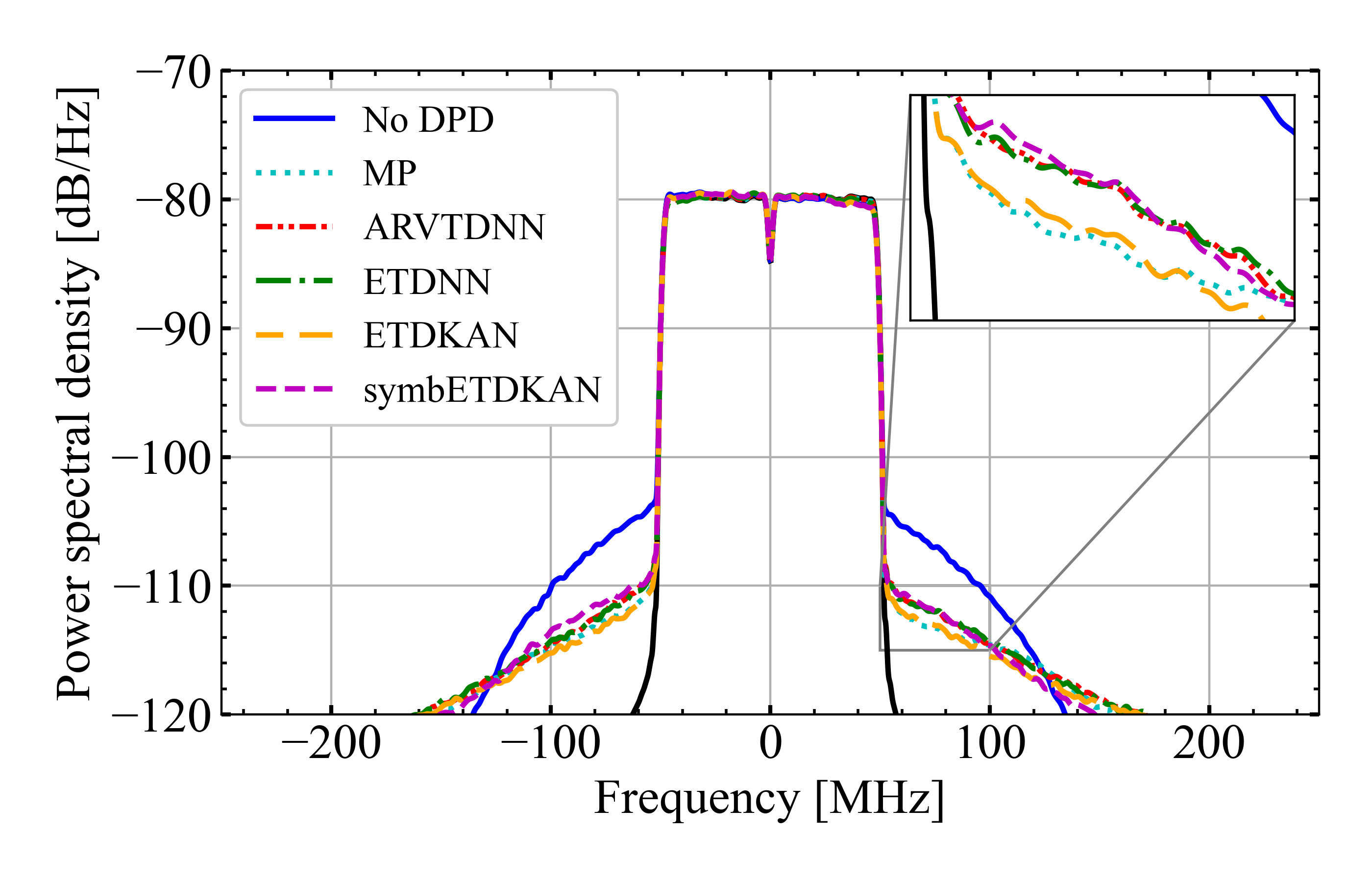}
    \caption{Spectrum of the received signals without and with DPD.}
    \label{fig:Rx_spec_exp}
\end{figure}

Fig.~\ref{fig:Rx_spec_exp} shows the received spectra without DPD and with the DPD models applied. A reduction in out-of-band emissions caused by nonlinear distortions can be observed, corresponding to an ACLR improvement of 5.62~dB for ETDKAN-DPD and 4.25~dB for symbETDKAN-DPD. These results further demonstrate that the proposed model can be successfully applied in an experimental environment with high nonlinearity.

\begin{table}[!t]
    \centering
    \caption{Performance metrics of the selected DPD models in experimental setup.}
    \begin{tabular}{|c | c | c | c|}
        \hline
        \textbf{DPD model} & \textbf{NFLOP} & \textbf{EVM [\%]} & \textbf{ACLR [dB]} \\   \hline
        No DPD & -- & 9.57 & -28.87 \\ \hline
        MP & 60 & 6.94 & -34.15 \\ \hline
        ARVTDNN & 852 & 7.61 & -33.60 \\ \hline
        ETDNN  & 620 & 7.42 & -33.63 \\ \hline
        ETDKAN & 5804 & 7.56 & -34.49 \\ \hline
        symbETDKAN & 53 -- 143 & 7.18 & -33.12 \\ \hline
    \end{tabular}
    \label{tab:dpd_metrics_exp}
\end{table}

\vspace{-0.25cm}
\section{Conclusion}
This article presents the first analysis of the application of KAN network–based DPD models to A-RoF systems. The ETDKAN-based DPD model was proposed and experimentally demonstrated for the compensation of nonlinearities in A-RoF links.

In the numerical analysis, the ETDKAN networks exhibited superior ACLR ($-36.54$ dB) performance compared to approaches based on MLP ($-35.62$ dB for ETDNN, and $-35.48$ dB for ARVTDNN);  however, this improvement was achieved at the cost of higher computational complexity, with NFLOP = 5804. By applying network symbolization, the NFLOP requirement of symbETDKAN was reduced to a level close to that of the MP model, between 53 and 143, that is the simplest among the evaluated approaches, while only marginally affecting performance (ACLR = $-35.71$ dB).

A back-to-back A-RoF experimental system was then employed to validate the applicability of the ETDKAN model in a practical environment. The results followed patterns similar to those observed in the numerical analysis, with the ETDKAN achieving the best performance of in therms of ACLR ($-34.49$ dB), and symbETDKAN achieving performance comparable to other NN-based models, with ACLR = $-33.12$ dB, while requiring significantly lower NFLOP. 

Future work could investigate the performance of KAN-based models under different nonlinear regimes, as well as assess the impact of the frequency-selective response of the optical channel with direct-detection. Due to their ability to derive analytical expressions for system representation, KAN networks have strong potential for application in the equalization and modeling of nonlinearities that are inherent in modern communication systems, including those used in 5G and 6G technologies.


%



\vspace{-0.25cm}
\section*{Acknowledgment}

This work was supported by the Programa de Pós-graduação em Engenharia Elétrica (PPGEE) of the Universidade Federal de Campina Grande (UFCG). This study has been partially funded by the Coordenação de Aperfeiçoamento de Pessoal de Nível Superior - Brasil (CAPES) - Finance Code 001, The National Council for Scientific and Technological Development (CNPq) - Grant 406684/2021-9, and the project PHOSENSE DEVICES supported by the EMBRAPII VIRTUS COMPETENCE CENTER IN INTELLIGENT HARDWARE FOR INDUSTRY - VIRTUS-CC, with financial resources from the HardwareBR PPI of the MCTI grant number 055/2023, signed with EMBRAPII.

\ifCLASSOPTIONcaptionsoff
  \newpage
\fi



\bibliographystyle{IEEEtran}
\vspace{-0.25cm}
\bibliography{IEEEabrv,bibliography}
\end{document}